\input harvmac.tex
\input epsf.tex
\newcount\figno
\figno=0
\def\fig#1#2#3{
\par\begingroup\parindent=0pt\leftskip=1cm\rightskip=1cm\parindent=0pt
\baselineskip=11pt
\global\advance\figno by 1
\midinsert
\epsfxsize=#3
\centerline{\epsfbox{#2}}
\vskip 12pt
{\bf Figure \the\figno:} #1\par
\endinsert\endgroup\par
}
\def\figlabel#1{\xdef#1{\the\figno}}
\def\encadremath#1{\vbox{\hrule\hbox{\vrule\kern8pt\vbox{\kern8pt
\hbox{$\displaystyle #1$}\kern8pt}
\kern8pt\vrule}\hrule}}

\batchmode
  \font\bbbfont=msbm10
\errorstopmode
\newif\ifamsf\amsftrue
\ifx\bbbfont\nullfont
  \amsffalse
\fi
\ifamsf
\def\IR{\hbox{\bbbfont R}}
\def\IZ{\hbox{\bbbfont Z}}
\def\IF{\hbox{\bbbfont F}}
\def\IP{\hbox{\bbbfont P}}
\else
\def\IR{\relax{\rm I\kern-.18em R}}
\def\IZ{\relax\ifmmode\hbox{Z\kern-.4em Z}\else{Z\kern-.4em Z}\fi}
\def\IF{\relax{\rm I\kern-.18em F}}
\def\IP{\relax{\rm I\kern-.18em P}}
\fi

\overfullrule=0pt

%


\lref\dmmv{ dmmv }
\lref\emss{
S.~Elitzur, G.~Moore, A.~Schwimmer and N.~Seiberg,
``Remarks On The Canonical Quantization Of The
Chern-Simons-Witten Theory,''
Nucl.\ Phys.\  {\bf B326}, 108 (1989).
}

\lref\cveticyoum{ M.~Cvetic and D.~Youm,
 ``General Rotating Five Dimensional Black Holes of Toroidally
Compactified Heterotic String,''
Nucl.\ Phys.\  {\bf B476}, 118 (1996)
[hep-th/9603100].
}

\lref\cveticlarsen{
M.~Cvetic and F.~Larsen,
``Near horizon geometry of rotating black holes in five dimensions,''
Nucl.\ Phys.\  {\bf B531}, 239 (1998)
[hep-th/9805097].
}
\lref\wadia{J.~R.~David, G.~Mandal, S.~Vaidya and S.~R.~Wadia,
``Point mass geometries, spectral flow and AdS(3)-CFT(2)
correspondence,''
Nucl.\ Phys.\  {\bf B564}, 128 (2000)
[hep-th/9906112].
}

\lref\giant{
J.~McGreevy, L.~Susskind and N.~Toumbas,
``Invasion of the giant gravitons from anti-de Sitter space,''
JHEP {\bf 0006}, 008 (2000)
[hep-th/0003075].
 }

\lref\long{
J.~Maldacena, J.~Michelson and A.~Strominger,
``Anti-de Sitter fragmentation,''
JHEP {\bf 9902}, 011 (1999)
[hep-th/9812073].
}
\lref\swlong{N.~Seiberg and E.~Witten,
``The D1/D5 system and singular CFT,''
JHEP {\bf 9904}, 017 (1999)
[hep-th/9903224].
}

\lref\maoz{
M.~Henneaux, L.~Maoz and A.~Schwimmer,
``Asymptotic dynamics and asymptotic
symmetries of three-dimensional  extended AdS supergravity,''
Annals Phys.\  {\bf 282}, 31 (2000)
[hep-th/9910013].
}
\lref\ms{J.~Maldacena and A.~Strominger,
``AdS(3) black holes and a stringy exclusion principle,''
JHEP {\bf 9812}, 005 (1998)
[hep-th/9804085].
}

\lref\vijay{
V.~Balasubramanian, J.~d.~Boer, E.~Keski-Vakkuri and S.~F.~Ross,
``Supersymmetric Conical Defects:
Towards a string theoretic description of black hole formation,''
[hep-th/0011217].
}

\lref\review{
O.~Aharony, S.~S.~Gubser, J.~Maldacena, H.~Ooguri and Y.~Oz,
``Large N field theories, string theory and gravity,''
Phys.\ Rept.\  {\bf 323}, 183 (2000)
[hep-th/9905111].
}

\lref\gianttwo{A.~Hashimoto, S.~Hirano and N.~Itzhaki,
``Large branes in AdS and their field theory dual,''
JHEP {\bf 0008}, 051 (2000)
[hep-th/0008016]. M.~T.~Grisaru, R.~C.~Myers and O.~Tafjord,
``SUSY and Goliath,''
JHEP {\bf 0008}, 040 (2000)
[hep-th/0008015]. S.~R.~Das, A.~Jevicki and S.~D.~Mathur,
``Vibration modes of giant gravitons,''
hep-th/0009019. S.~R.~Das, A.~Jevicki and S.~D.~Mathur,
``Giant gravitons, BPS bounds and noncommutativity,''
hep-th/0008088. S.~R.~Das, S.~P.~Trivedi and S.~Vaidya,
``Magnetic moments of branes and giant gravitons,''
JHEP {\bf 0010}, 037 (2000)
[hep-th/0008203]. A.~Mikhailov,
``Giant gravitons from holomorphic surfaces,''
JHEP {\bf 0011}, 027 (2000)
[hep-th/0010206].
}

\lref\giantthree{M.~T.~Grisaru, R.~C.~Myers and O.~Tafjord,
JHEP {\bf 0008}, 040 (2000)
[hep-th/0008015].
}
\lref\giantfour{A.~Mikhailov,
JHEP {\bf 0011}, 027 (2000)
[hep-th/0010206].
}
\lref\giantfive{S.~R.~Das, A.~Jevicki and S.~D.~Mathur,
[hep-th/0009019].
}
\lref\giantsix{S.~R.~Das, A.~Jevicki and S.~D.~Mathur,
[hep-th/0008088].
}

\lref\giantseven{
S.~R.~Das, S.~P.~Trivedi and S.~Vaidya,
JHEP {\bf 0010}, 037 (2000)
[hep-th/0008203].
}
\lref\gianteigth{
J.~Lee,
[hep-th/0010191].
}

\lref\chring{
W.~Lerche, C.~Vafa, N.P.~Warner,
``Chiral Rings in N=2 Superconformal Theories,''
Nucl.\ Phys.\ {\bf B324}, 427 (1989).
}

\lref\chone{O.~Coussaert and M.~Henneaux,
``Supersymmetry of the (2+1) black holes,''
Phys.\ Rev.\ Lett.\  {\bf 72}, 183 (1994)
[hep-th/9310194].
}

\lref\mms{
J.~Maldacena, G.~Moore and A.~Strominger,
``Counting BPS black holes in toroidal type II string theory,''
[hep-th/9903163].
}

\lref\wjones{
E.~Witten,
``Quantum Field Theory And The Jones Polynomial,''
Commun.\ Math.\ Phys.\  {\bf 121}, 351 (1989).
}

\lref\wadia{
J.~R.~David, G.~Mandal, S.~Vaidya and S.~R.~Wadia,
``Point mass geometries, spectral flow and AdS(3)-CFT(2) correspondence,''
Nucl.\ Phys.\  {\bf B564} (2000) 128
[hep-th/9906112].
}
\lref\ti{
J.~M.~Izquierdo and P.~K.~Townsend,
``Supersymmetric space-times in (2+1) adS supergravity models,''
Class.\ Quant.\ Grav.\  {\bf 12}, 895 (1995)
[gr-qc/9501018].
}

\lref\db{
J.~de Boer,
``Six-dimensional supergravity on S**3 x AdS(3) and 2d conformal field  theory,''
Nucl.\ Phys.\  {\bf B548}, 139 (1999)
[hep-th/9806104].
}

\lref\tseytlin{
M.~Cvetic, A~.Tseytlin,``Sigma model of near extreme rotating black holes
and their microstates,'' Nucl.\ Phys.\ {\bf B537}, 381 (1999)
[hep-th/9806141]. A.~ Tseytlin, `` Extreme dyonic black holes in string
theory'', Mod.\ Phys.\ Lett.\ {\bf A11}, 689 (1996)
[hep-th/9601177]. A.~ Tseytin, ``Generalized chiral null models and
rotating string backgrounds'', Phys.\ Lett.\ {\bf B381}, 73 (1996)
[hep-th/9603099].
}


\Title{\vbox{\baselineskip12pt
\hbox{\tt HUTP-00/A049}
\hbox{\tt IHP-2000/08}
\hbox{\tt hep-th/0012025}}}
{\vbox{\centerline{De-singularization by rotation}
}}
\bigskip
\centerline{ Juan Maldacena$^{1,2}$ Liat Maoz$^1$ }
\bigskip
\centerline{$^1$ Jefferson Physical Laboratory}
\centerline{Harvard University}
\centerline{Cambridge, MA 02138, USA}
\bigskip
\centerline{$^2$ Institute for Advanced Study}
\centerline{Princeton, NJ 08540}

\vskip .3in

We consider certain BPS supergravity solutions of string theory which
have singularities  and
we show that the singularity goes away when we add angular momentum.
These smooth solutions enable us to obtain {\it global } $AdS_3$ as the
near horizon geometry of a BPS brane system in an asymptotically flat
space.


\newsec{Introduction}

It is well known that some BPS states in four and five
dimensional supergravity theories can be realized as
non-singular extremal black holes with non-zero horizon area.
This is the situation for generic black hole charges. However,
there
are  some cases  where the area of the horizon becomes
zero and the geometry becomes singular. For
example, this happens for 1/4 BPS states of string theory on $T^5$.
In this paper we show that by considering  1/4 BPS states with maximal
angular  momentum we can produce
a completely non-singular geometry once we suitably include
one of the internal dimensions.
We were led to this solution by thinking about supersymmetric
conical singularities in $AdS_3$.
So first we
analyze various aspects of supersymmetric $AdS_3$
spaces and conical singularities \ti .
When we are dealing with $AdS_3$ we can consider the theory with
NS-NS or RR boundary conditions on the spatial circle.
It known that the $M=0$ BTZ black hole is a RR ground state \chone .
We show that by introducing Wilson lines for $U(1)$ gauge fields
in $AdS_3$ we can also interpret other conical singularities as
RR ground states. Even pure $AdS_3$ with a suitable Wilson line
can be interpreted as a RR ground state. All these ground states
are different in their $U(1)$ charges.

If we view  global $AdS_3$ as the near horizon region of a
six  dimensional  rotating
black string  of string theory on $ R^{1,4} \times
S^1 \times M^4 $
coming from D1-D5 branes  wrapped on $S^1$, \foot{
The D5 branes also wrap $M^4$.} then we can match the
smooth {\it global} $AdS_3$ solution to asymptotically flat space in such
a way that it preserves supersymmetry. In other words, by
adding angular momentum we can find a smooth supergravity
solution that corresponds to the D1-D5 system. These are solutions
which have maximal angular momenta $ J_L = \pm J_R =
Q_1Q_5/2 \equiv k/2  $. \foot{ One can also view the system in the S-dual
picture, involving $F1-NS5$. Conformal models describing the $F1-NS5$
system with couplings representing the angular momenta have been
discussed in \tseytlin .}

These $AdS_3$ geometries with Wilson lines can also have the
interpretation of ``giant gravitons'' in $AdS_3$.


The proper interpretation of these solutions will involve a
precise statement and understanding of the possible boundary
conditions for the gauge fields that live on $AdS_3$. So in
section 2 we review some facts about gauge fields
 and  Chern Simons theory.
In section 3 we describe the interpretation of the solutions
from the AdS/CFT point of view.
In section 4 we match the $AdS_3$ solutions to the  asymptotically
flat region. In section 5 we briefly remark about the interpretation
of these configurations as giant gravitons.

As this paper was in preparation we received \vijay\ which
has a great deal of overlap with this paper.

\newsec{ Some facts about Wilson lines and Chern Simons theory }

Let us start by describing some facts about $U(1)$ gauge fields.
Suppose we have a plane described by coordinates $\rho , \varphi$,
$ds^2 = d\rho^2 + \rho^2 d\varphi$. Then consider a gauge
field with the connection $ A_\varphi = a $ where $a$ is any constant.
We see that $F =0$ everywhere in the plane except at the origin
where it is a delta function. This is of course the familiar gauge
field of a Bohm-Aharonov vortex.
The interaction
with the gauge field is normalized so that we get the phase
$e^ {i \int A} $ for the field with the minimal quantum of charge.
 We see that if $a$ is an integer particles do not feel any field
and indeed we can set $A$ to zero by a gauge
transformations $A \to A + d \epsilon$ where
 $ \epsilon(\varphi + 2 \pi) = \epsilon(\varphi) + 2 \pi n $, with $n$
integer.
We need to specify the boundary conditions for the charged
fields when we go
around the origin. We will work with fixed boundary conditions
for the fields and we will vary $a$.
Suppose we have a fermionic field and we impose the boundary
condition that $\psi$ is periodic as it goes around the circle.
Then if we set  $a=1/2$, the field  will effectively become
antiperiodic. This  implies that the fermionic field
will be totally continuous at the origin, since the
minus sign is what we expect for a rotation by $2 \pi$.

Now let us suppose that we have Chern Simons theory on
a solid cylinder $D_2\times R$, where $D_2$ is a disk.
 Then we need to impose some
boundary conditions on the gauge field.
As shown in \wjones \emss\  we can
impose the boundary condition only on
one component of $A$ along the boundary. One way to understand this
is to view  the direction orthogonal to the boundary as time  so
that one realizes that the two components of $A$ along the boundary
are canonically conjugate variables.
We will be interested in setting boundary conditions of the form
$2 A_- = A_0 - A_\varphi =0$.
 It is easy to see that these boundary conditions
are consistent. We choose these boundary conditions because it was
shown in \maoz\ that they are appropriate for gauge fields in $AdS$
supergravities.
Once we give these boundary conditions we can have a variety of
states  in the theory with various values of $A_+$ on the boundary.
These values are $2A_+ = q/2 k$, with $q$ integer \emss , and $k$ the
level of the Chern Simons theory. These various
states can arise by inserting various Wilson lines in the interior.
States with $ q \to q \pm 2 k$ are related by a large gauge transformation
which does not vanish at the boundary. These transformations
map physical states to other physical states in the boundary theory.
 {} From the point of view of the topological theory in the bulk,
states with $ q$ and $q \pm 2k $ are equivalent.
%
The $U(1)$ charge of the state has the value
of ${ 1\over 2 \pi } \int A $ along the spatial circle.
If we have a Wilson line of charge $q$
 in the interior, this value is  $A_\varphi =
q/(2 k) $.

Similar remarks about CS theory apply when the gauge group is
non-compact, such as $SL(2,R)$.
In this case we consider again configurations with vanishing
field strength and with the same asymptotic boundary conditions.
This implies that the space is locally $AdS$ but not globally.
For example, we can consider the  conical space

\eqn\coni{ ds^2 = - ( r^2 + \gamma^2)
dt^2 + r^2 d\varphi^2 + { dr^2 \over
r^2 + \gamma^2}
}

Locally this is an $AdS_3$ space, but at $ r=0$ we have a conical
singularity if $\gamma \not =1 $.






\newsec{Conical singularities and AdS/CFT}

In this section we will apply some of the above remarks to
supergravity theories on $AdS_3$. What we will describe
is mainly contained in \ti \wadia .
We will consider supergravity
theories with extra $U(1)$
 gauge fields on $AdS_3$.
One example we have in mind is the case of string theory on
$AdS_3 \times S^3 \times K3$, but other examples could be treated
in a similar way. We will consider gravity theories on
$AdS_3$ with at least $(2,2)$ supersymmetry. This implies that
we will have $U(1)_L \times U(1)_R$ gauge fields.
Pure three dimensional gravity on
$AdS_3$ is given by an $SL(2,R)^2$ Chern Simons
theory, which we will use to describe the conical spaces.
In this situation we could consider solutions with arbitrary Wilson
lines for  the $U(1)_{L,R}$ gauge field as well as the
$SL(2,R)_{L,R} $ gauge
fields. In principle these solutions are singular in the interior
and we should not consider them, unless we have a good reason to
think that the singularity will be resolved in the full theory.

In this paper we will consider  singularities which preserve
at least (2,2) supersymmetry.
We will impose RR boundary conditions on the
fields and we consider arbitrary Wilson lines. In order for the
solution to be supersymmetric the Wilson line in the $SL(2,R)$ part and
the $U(1)$ part should be essentially the same, we will later
make this statement more precise. The boundary of $AdS_3$ is
$R \times S^1$. We normalize charges
so that a fermion carries integer charge under $U(1)_{R,L}$.
As standard in AdS/CFT,  the boundary conditions on all supergravity
fields correspond to the microscopic definition of the ``Lagrangian''
of the CFT, including the periodicities of the fields as
we go around the circle, etc.
 We can then consider all solutions to the supergravity equations
 with given
boundary conditions. Different solutions correspond to different
states in the boundary CFT.
Now let us choose RR boundary conditions for the  CFT  and sugra
fields on  the
spatial boundary circle.
We will impose the boundary
condition $A^L_- = A^R_+= 0$ for $U(1)_{L,R}$.\foot{Actually one could
impose the boundary condition $A_-^L=\epsilon_L~,~A_+^R=\epsilon_R$, where
$\epsilon_{L,R}$ are some constants. These would correspond to left and
right spectral flows with the parameters $\epsilon_{L,R}$.}

We will consider flat gauge fields with $U(1)_{R,L}$ connections
given by constant values $A^L_+ = a_+ $, $A^R_- = a_- $.
Supersymmetry determines the three dimensional geometry.
We consider spinors generating supersymmetry that are periodic
when we go around the circle, since we said we are interested in
the RR sector.
The solution is then:\foot{We use conventions where $R_{AdS}=1$. }
\eqn\solgen{
\eqalign{
{{ds^2_3} } =& -[(r - {{a^2_+ - a^2_-} \over r} )^2 +4a^2_+] dt^2 +
 { {dr^2} \over {(r - { {a^2_+-a^2_-} \over r} )^2 +4a^2_+} } + \cr
&+ r^2 (d\varphi + {{a^2_+-a^2_-} \over r^2} dt)^2 \cr
& A^L_+ = a_+ ~~~~, ~~~~ A^R_-=a_- ~~~~, ~~~~~~~A^L_-=A^R_+=0
}}

In the particular case of $a_+ = a_- = \gamma/2$ the solution
is

\eqn\sol{\eqalign{
{ {ds^2_3} } = &- (r^2 +  \gamma^2) dt^2 + r^2 d\varphi^2  +
{dr^2 \over r^2 + \gamma^2}
\cr
 & A^L_+ = A^R_- = \gamma/2 ~~~~, ~~~~~~~A^L_-=A^R_+=0
}}

All these configurations have zero energy, as implied by the
RR sector super-algebra. The $AdS_3$ space in \sol\ seems to have
negative energy, but one should add to this the energy
that comes from the Wilson line. This additional energy
comes from the ``singleton'' that lives at the boundary of
$AdS$ which encodes this degree of freedom. This was explicitly
shown in \maoz . So we have  $L_0 = \bar L_0 =0$.
 The angular momenta are half the $U(1)$ charges,
$J_L = J_R =  k \gamma /2 $.
So we see that $\gamma$ should be
quantized as $\gamma = n/k$.
We get zero energy states
with various amounts of angular momenta.

So what is the interpretation of these spaces? which ones are allowed
and which ones are not?. All these are supersymmetric solutions.
Almost all of them are singular. Only if  $\gamma =1$  we see
 from \sol\ that we get a nonsingular solution. Let us
discuss this solution first. The three dimensional geometry is that
of $AdS_3$. The Wilson line around the origin of $AdS_3$ is such
that it effectively
changes the periodicity of fermionic fields from periodic
to anti-periodic, so that they are smooth at  the origin.
This solution has angular momenta $J_L = J_R = k/2$. What is this
state in the boundary CFT?. We know that the boundary CFT has
a large number of RR vacua \review . These vacua have angular
momenta $|J_{L,R}| \leq k/2$. We see that the non-singular solution
corresponds to a state with the maximal value of the angular momentum.
 {} From general arguments \chring we know that there is
a single RR state with maximal value of the RR charge, it is the
state that maps to the NS vacuum under spectral flow. Here we
indeed see that the state we find is essentially the same as global
$AdS_3$ which was identified as the NS vacuum. The only difference
is that the Wilson lines  imply that particle energies are
shifted as they  are shifted  under spectral flow.
Now we turn to the solutions with $\gamma \not =1$.
All those solutions contain a singularity at the origin.
It is clear that starting from the solution with $\gamma = 1$ we
can add supergravity particles that decrease the angular
momentum and leave $L_0 = \bar L_0 = 0$, these particles, are of course,
the chiral primaries discussed in \ms , see also \db .
If we have particles
with high values of the angular momentum, $ l \gg 1$,  $ l/k $ fixed,
they will appear like very massive particles from the $AdS_3$ point
of view and will give rise to the conical spaces with $\gamma<1$.
It is not possible to get the conical spaces with $\gamma >1$
in this fashion, since all those  supergravity particles would increase
the energy and will remove us from the RR vacuum.
In other words, by adding supergravity particles  to the state
 with $J_L=J_R =k/2$ we can decrease the angular momentum while
preserving the zero energy condition. If we try to increase
$J$ we would increase the energy.

We could imagine decreasing $J$ by adding supergravity particles with
low values of the spin, those gravity particles have wavefunctions
which are quite extended in $AdS$. If we added them
in a coherent state, we should be able to find classical
solutions which are also smooth and do not have these conical
singularities. Finding these solutions would require us to use
the full six dimensional gravity equations.
In other words, the fact that for $J< k/2$ we only found singular
solutions does not mean that there are no non-singular solutions.
A trivial  example is  the following. Consider
the $AdS_3 \times S^3 $ case. Now we have $SU(2)_L$ and $SU(2)_R$
symmetry groups. Let us pick the $U(1)$'s in the above discussion
to be in the direction $\hat 3$. Take the
solution with maximal angular momentum and perform an $SU(2)_{L,R}$
 rotation
in the $\hat 1 $ axis so that now the angular momentum points
in the $\hat 2 $ direction. We get exactly the same
$AdS_3$ space but now with a Wilson line $A^{L,2}_+ =A^{R,2}_-=
 1/2$ and the rest
zero. This is a solution
with zero $U(1)$ charges but with no singularity,
as opposed to the solution in \sol\ with $\gamma =0$.
Of course here we are treating these Wilson lines in a classical
fashion. This is correct in the large $k$ limit where
we deal with macroscopic amounts of angular momentum.

It is easy to see that any solution which is $AdS_3$ and a
Wilson line of the form $A^L_+ = 1/2 +n $, $A^R_- = 1/2 + n'$ with
integer $n,n'$ will be non-singular. These solutions correspond to
the spectral flow of the  state with $n=n'=0$. These solutions do
not preserve the supersymmetries that the RR ground state preserve,
but they do preserve other supersymmetries. These are the
configurations that are related by spectral flow to the NS sector
ground state.

\fig{ Spectrum of the theory in the RR sector. RR
ground states have spins $|J|\leq k/2$. Quantum numbers  that lie within
the shaded region, with $L_0 > J^2/k$ can be carried by  black holes.
We have a similar figure for $\bar L_0$ and $\bar
J$.
}{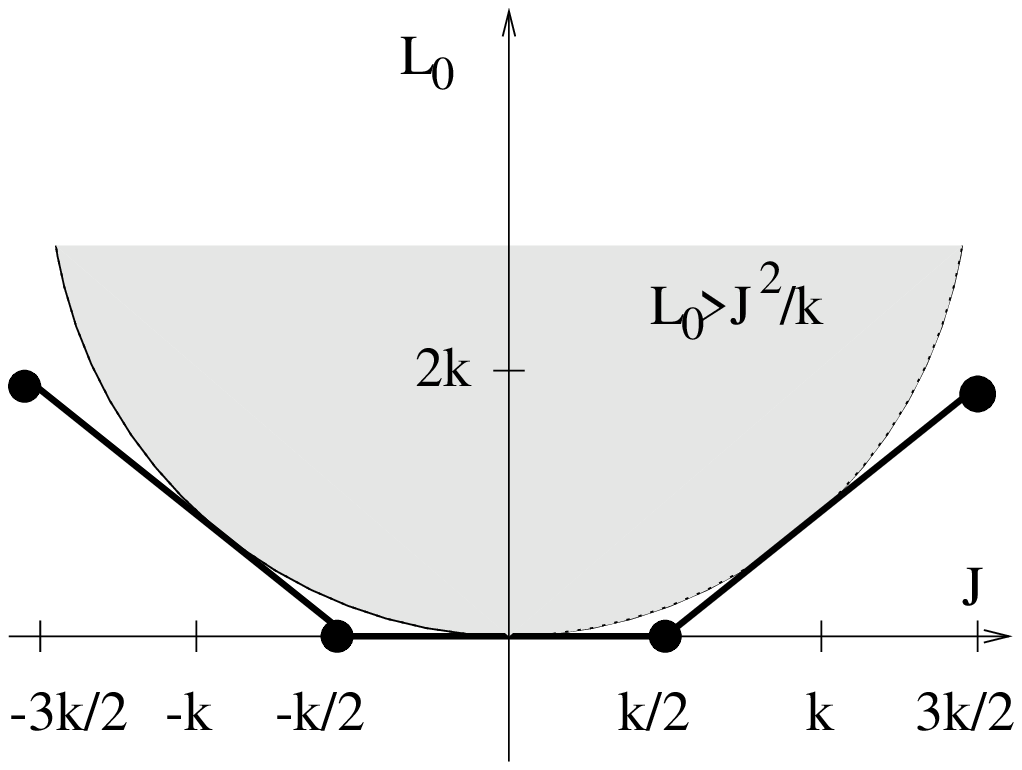}{4.5 truecm}
\figlabel\rsector

\fig{ Spectrum of the theory in the NS sector sector.
 Quantum numbers  that lie within the shaded region, with $L_0 > J^2/k +
k/4 $ can be carried by  black holes. States with $J = n k$, $L_0 =
n^2 k $ are $AdS_3$ spaces with  Wilson
lines.
}{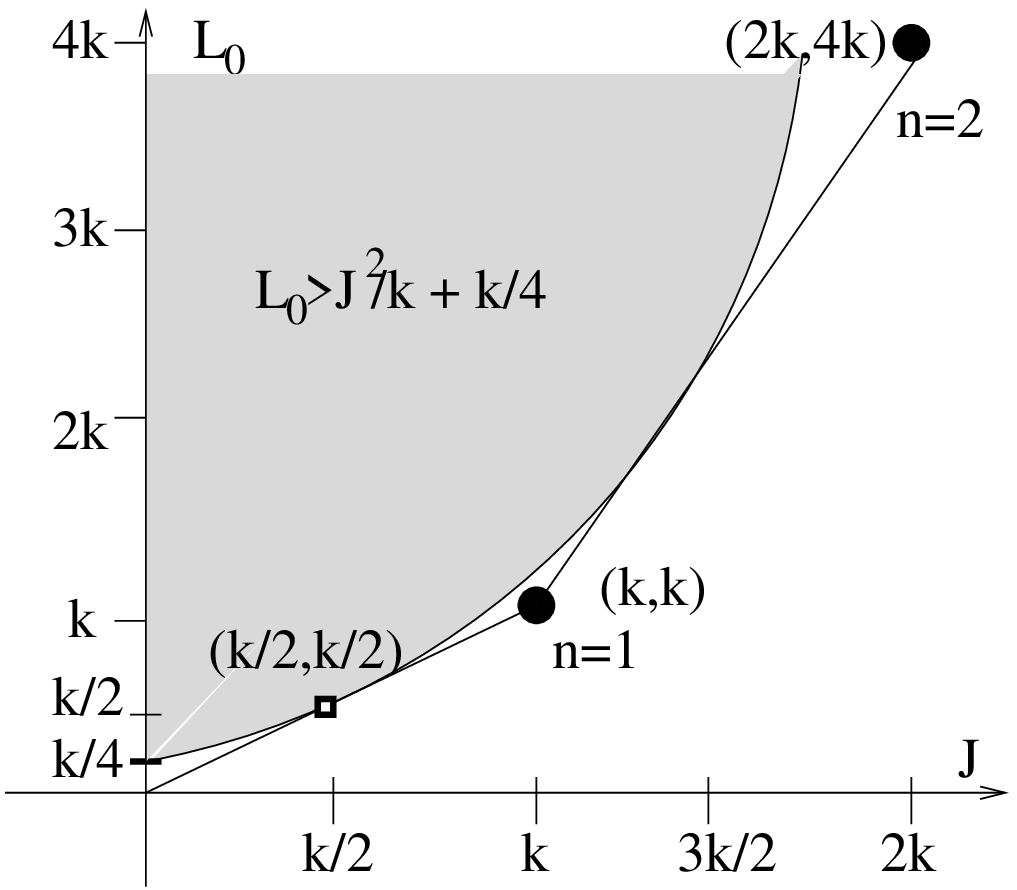}{5.00 truecm}
\figlabel\nssector

\newsec{Non-singular solutions in asymptotically flat space}

In this section we point out that  the conical spaces, including
the
non-singular $AdS_3$ space with a Wilson line, can be extended
to supersymmetric
 solutions of six dimensional supergravity that are asymptotic
to $R^5 \times S^1$, with periodic boundary conditions on $S^1$.
In other words, they represent BPS solutions in this six dimensional
string theory.

We can find the solution by starting with the most general five
dimensional black hole solution written in \cveticyoum , lifting
it up to six dimensions as in \cveticlarsen\ and taking the extremal
limit with zero momentum charge while keeping the angular momenta
nonzero.
The solution we obtain is parametrized by two angular momentum
parameters which we take
as $\gamma_{1,2}$:  $J_{L,R}={k \over 2}(\gamma_1
\mp \gamma_2)$ and can be written in the form\foot{
To relate our parameters and coordinates to the ones in
eq. (4) of Cvetic and Larsen \cveticlarsen, we have
$\gamma_{1,2}={R_y \over \sqrt{k}}(\cosh\delta_0\ell_{1,2}
-\sinh\delta_0\ell_{2,1})$ , $k=\lambda^4$ and $r={\sqrt{k} \over
R_y}r^{C.L.}$,  $t={1 \over R_y}t^{C.L.}$ , $\varphi={1 \over
R_y}y^{C.L.}$}:

\eqn\solsixgen{\eqalign{
{ds_6^2 \over \sqrt{k}} =& {1 \over h}(-dt^2+d\varphi^2) +
 hf(d\theta^2 + {{r^2 dr^2} \over {(r^2 +\gamma_1^2)(r^2 +\gamma_2^2)}})
\cr -& {2 \over {hf}} [(\gamma_2 dt +\gamma_1
d\varphi)\cos^2\theta d\psi + (\gamma_1 dt +\gamma_2
d\varphi)\sin^2\theta d\phi] + \cr +& h[(r^2 +\gamma_2^2)
+(\gamma_1^2-\gamma_2^2){{\cos^2\theta} \over
{h^2f^2}}]\cos^2\theta d\psi^2 + \cr +& h[(r^2 +\gamma_1^2)
-(\gamma_1^2-\gamma_2^2){{\sin^2\theta} \over
{h^2f^2}}]\sin^2\theta d\phi^2 \cr}} where: \eqn\defs{ \eqalign{ f
&= f(r,\theta) \equiv r^2 + \gamma_1^2 \cos^2 \theta +\gamma_2^2
\sin^2 \theta \cr h &= h(r,\theta) \equiv {\sqrt{k} \over R_y^2} (
1+{{R_y^2Q_1} \over {kf}})^{1/2}(1+{{R_y^2Q_5} \over {kf}})^{1/2}
\cr} } and $R_y$ is the radius of the $S^1$ parameterized by
$\varphi$.

Setting the two angular momenta equal ($\gamma_2=0~~,~~\gamma \equiv
\gamma_1$): $J_L=J_R=k\gamma /2$ , we get the solution:

\eqn\solsixcons{
\eqalign{
{{ds^2_6} \over \sqrt{k}} =& -{1 \over h}
(dt+{{\gamma\sin^2\theta} \over {r^2+\gamma^2\cos^2\theta}}d\phi )^2
+ {1 \over h}
(d\varphi - {{\gamma\cos^2\theta} \over
{r^2+\gamma^2\cos^2\theta}}d\psi)^2  + \cr
&+ h {{r^2+\gamma^2\cos^2\theta}\over{r^2+\gamma^2}}dr^2  +\cr
&+ h[(r^2+\gamma^2\cos^2\theta) d\theta^2 +(r^2+\gamma^2)\sin^2\theta
d\phi^2+r^2\cos^2\theta d\psi^2] \cr
}}

In the decoupling near-horizon limit the metric reduces to a locally
$AdS_3 \times S^3$, where the $S^3$ angles are defined as
$\tilde{\psi}=\psi-\gamma\varphi~,~\tilde{\phi}=\phi-\gamma t$
\cveticlarsen .

Since the original angles are identified as:
$\varphi \sim \varphi + 2\pi$ , $\theta \sim \theta + \pi/2$, $\psi \sim
\psi + 2\pi$ , $\phi \sim \phi +2\pi$, these new coordinates have the
global identifications:

\eqn\ide{ \eqalign{
(\varphi,\tilde{\psi}) &\sim (\varphi, \tilde{\psi})+
2\pi (1,-\gamma ) \sim (\varphi,\tilde{\psi})+2\pi (0,1) \cr
\theta &\sim \theta +\pi /2 \cr
\tilde{\phi} &\sim \tilde{\phi} +2\pi \cr}
}

For general (noninteger) values of the parameter $\gamma$ the
periodicities of the $AdS_3$ and the $S^3$ parts are still coupled, and
the geometry obtained is singular.

The most interesting solution is the one with angular momenta
$J_L = J_R = k/2$, when $\gamma=1$, since it is non-singular. It is a
non-singular, geodesically complete geometry.
In its decoupling  near-horizon limit, the space is globally a direct
product $AdS_3 \times S^3$, as can be seen looking at the periodicities of
the angles in \ide.

It seems that the fact that this solution is non-singular is
related to the fact that there are very few states in the CFT
with similar values of the angular momenta.

It would be interesting to see if other BPS states in the $AdS_3$
region could be  matched to the asymptotically flat region.
Natural candidates are states with $L_0 =0, J^R = k/2 $ and
$J_L = k/2 + nk $. In the near horizon limit these states have
$A^R_- = 1/2$, $A^L_+ = 1/2 + n$. The elliptic genus formula
tells us that there is a single BPS state with these values of
the angular momenta\foot{ In principle, we also need to add
the center
of mass motion of the string in the transverse four dimensions
\mms .} (it is just the left spectral flow of the
state we found above). If we tried to take a  limit of the
solutions in \cveticyoum \cveticlarsen , we would find
that $J_R =0$. It could be that we need to make a more general
ansatz.

\newsec{Super giant gravitons}

In this section we consider NS-NS boundary conditions on the circle
at the boundary. The ground state is $AdS_3$ with no Wilson lines
for  the $U(1)$ gauge fields. We can consider the spectrum of chiral
primaries, i.e. states with $L_0 = J_L$, $\bar L_0 = J_R$ as in \ms .
>From the CFT point of view we can calculate how many of these states
we expect. It turns out that there is a single state with
$J_{L,R} =0 $ and the number of states increases as we increase
the values of $J_{L,R}$, it reaches a maximum at
$J_L = J_R  = k/2 $ and then it starts decreasing again so that
for
$J_L = J_R =  k $ we find just a single state again.
In other AdS compactifications there is a  maximum value
for the single particle BPS states. In \giant\  it was shown that
these states are realized as expanded branes, see also \gianttwo  .
In $AdS_5 \times S^5 $
the cutoff appears at $ J = N$ \review ,
 where $J$ is the angular momentum on
$S^5$. In $AdS_3$ the situation is  different \ms ,  there
is an absolute cutoff on $J$ at $J_L = J_R =  k$, there are no
chiral primary  states beyond this value of $J$.
By using the previous ideas about Wilson lines it is easy to see
that this state is just $AdS_3$ with $U(1)_L \times U(1)_R$ Wilson
lines equal to $A^L_+ = A^R_- = 1$.
We could roughly think about it as an $AdS$ space which is just
rotating as a whole. Only the
``singleton'' field is excited.
 The singleton is the mode that
appears at the boundary from the Chern Simons theory in the interior.
We can say that  gravitons became so big that they
live at the boundary of $AdS$.

So in the $AdS_3\times S^3$ case the graviton with maximum
angular momentum is not an expanded brane  but just a
different classical solution. This is in agreement with the fact
that the maximal spin, $k$, is of the order of
the inverse six dimensional Newton's constant, while in the
$AdS_5 \times S^5$ this maximal value, $N$, is proportional to the
square root of Newton's constant.
Notice that objects such as long strings \long \swlong\  are not
of concern here since we can work at a point in moduli space
where there is no finite energy long string at infinity. This
is possible if $Q_1$ and $Q_5$ are coprime \swlong .

In summary, as we pile up chiral primary particles on $AdS_3$ we
get to a point at $J_L = J_R = k/2$ where we are on the verge
of making a black hole \review . If $J_L , J_R $ approach their maximal
values we have again a smooth geometry with a small number
(if $J_{L,R}$
 are sufficiently close to $k$) of chiral primary particles.

It would be interesting to see if something similar happens for
other $AdS$ spaces.

{\bf Acknowledgments}

We  would like to thank G. Gibbons, G. Moore, C. Pope, A. Schwimmer, N.
Seiberg, and A. Strominger for discussions. LM would also like to thank
the Institut Henri Poincare for hospitality. We would like to
thank A. Tseytlin for pointing out  \tseytlin .

This  research
was supported in part by DOE grant DE-FGO2-91ER40654,
NSF grant PHY-9513835, the Sloan Foundation and the
David and Lucile Packard Foundation.

\listrefs

\bye